\documentclass[aps,pra,twocolumn,amsmath,amssymb,nofootinbib,showpacs,superscriptaddress,longbibliography]{revtex4-1}
\usepackage[english]{babel}
\usepackage{latexsym}
\usepackage{graphics}
\usepackage{graphicx}
\usepackage{epsfig}
\usepackage{color}
\usepackage{bm}
\usepackage{amsmath}
\usepackage{amssymb}
\usepackage{amsthm}
\usepackage{dcolumn}
\usepackage{bm}
\usepackage{float}
\usepackage{hyperref}
\usepackage{color}
\usepackage{epstopdf}
\usepackage{braket}
\usepackage{cleveref}
\usepackage[svgnames]{xcolor}
\hypersetup{hidelinks,colorlinks=true,allcolors=DarkBlue}
\usepackage{tikz}
\usepackage[toc,page]{appendix}
\usetikzlibrary{quantikz2}

\theoremstyle{statement}

\begin{document}

\preprint{APS/123-QED}

\title{All-to-all connectivity of Rydberg-atom-based quantum processors \\  with messenger qubits}

\author{Ivan V.                                  Dudinets}
\affiliation{Russian Quantum Center, Skolkovo, Moscow 121205, Russia}
\affiliation{Moscow Institute of Physics and Technology, Institutskii per. 9, Dolgoprudnyi, 141700, Russia}
\author{Stanislav S. Straupe}
\affiliation{Russian Quantum Center, Skolkovo, Moscow 121205, Russia}
\affiliation{Quantum Technology Centre and Faculty of Physics, M. V. Lomonosov Moscow State University, Moscow 119991, Russia}
\author{Aleksey K. Fedorov}
\affiliation{Russian Quantum Center, Skolkovo, Moscow 121205, Russia}
\affiliation{National University of Science and Technology ``MISIS'', Moscow 119049, Russia}
\author{Oleg V. Lychkovskiy}
\affiliation{Skolkovo Institute of Science and Technology, Moscow 121205, Russia}
\affiliation{Department of Mathematical Methods for Quantum Technologies,\\
		Steklov Mathematical Institute of Russian Academy of Sciences\\
		8 Gubkina St., Moscow 119991, Russia
}	
\affiliation{Russian Quantum Center, Skolkovo, Moscow 121205, Russia}

\date{\today}
\begin{abstract}
Rydberg atom arrays are a front-running platform for quantum processors.  A major challenge threatening the scalability of this platform is the limited qubit connectivity due to the finite range of interatomic interactions. We explore an approach to realize dynamical all-to-all connectivity with the use of moving ``messenger’’ atomic qubits that couple distant ``computational’’ qubits held in a static tweezer array. We detail and compare five specific architectures based on this concept, each presenting distinct advantages and challenges tied to the efficacy of techniques used to couple, move and measure atomic qubits. We demonstrate that, though technologically demanding, the messenger-qubit paradigm opens a promising avenue to a truly scalable quantum processor based on Rydberg atoms.
\end{abstract}

\maketitle


\section{Introduction}

Neutral atoms are promising candidates for the role of qubits~\cite{Saffman2010} and qudits~\cite{Kiktenko2023}, which are elementary building blocks of quantum computational devices~\cite{Brassard1998, Ladd2010, Fedorov2022}. 
Major  advantages of neutral atoms include inherent indistinguishability, long coherence times and the the ability to control large-scale atomic ensembles by optical means~\cite{Lukin2014-2}. 
Neutral atoms can be trapped and individually controlled with optical tweezers, allowing the assembly of atoms in defect-free arrays~\cite{Browaeys2016, Browaeys2016-2, Lukin2016}.
A downside of atomic qubits is the fact that atoms almost do not interact with each other under normal conditions, while the interaction is an essential prerequisite for the exchange of quantum information~\cite{Browaeys2020-2}. A nowadays standard way to ensure strong interatomic interactions is to excite  atoms to Rydberg states~\cite{Lukin2000, Lukin2001,Saffman2010}. 
This approach has been actively used to develop large-scale quantum computational devices~\cite{Browaeys2016,Browaeys2016-2,Lukin2016,Zoller2017,Lukin2017,Lukin2019,Browaeys2020-2,Lukin2021-7,zkpl-hh28}, which are now about to compete with the most powerful classical supercomputers in solving problems of physics simulation~\cite{bernien2017probing,Browaeys2019,Browaeys2020,Lukin2021,Lukin2021-2,Saffman2022,zhang2025probing} and optimization problems~\cite{Lukin2021-6,Lukin2022,Lukin2023,de2025demonstration,bombieri2025quantum}. 

Yet,  interactions are strong only between nearby Rydberg atoms, with their strength abruptly falling  with distance. The local nature of two-qubit coupling is a fundamental challenge plaguing not only atomic arrays but, to varying degrees, most quantum computational platforms --- including superconducting circuits \cite{Bravyi_2022_Future_of_quantum_sc_computers} and quantum dots \cite{Zhang_2018_Semiconductor_quantum_computation_review}.
This constraint threatens the practical realization of universal quantum computation, which requires on-demand pairwise interactions between any two qubits in a processor.

\begin{figure}[t] 
\centering
\includegraphics[width=0.7\linewidth]{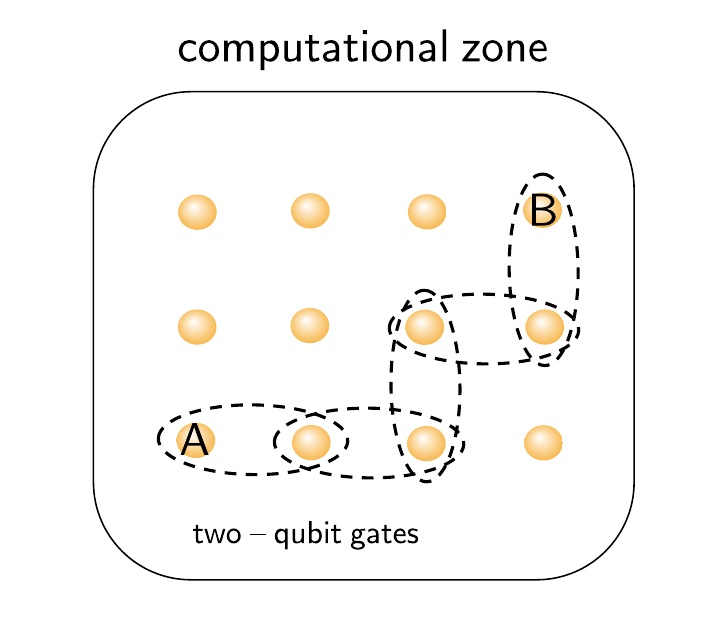}
\caption{A straightforward way to perform a logic gate between  two distant target qubits,  $A$ and $B$,  is to perform a sequence of physical gates between neighbouring qubits along some path. The drawback of this approach  is the growth of the physical gate count per a logic gate  with the processor size and, as a consequence,  the exponential drop of the logic gate fidelity with the number of qubits.}
\label{fig:neighboring}
\end{figure}

A common approach to realize all-to-all connectivity is to promote local {\it physical}  connectivity to the all-to-all {\it logic} one at the expense of gate count. Namely, to perform a two-qubit logic gate between two distant target qubits, one performs a sequence of physical gates, each involving only neighboring qubits, along a path connecting two target qubits \cite{Martinis2019,Jeong_2022_Rydberg,sun2024buffer, he2024distant,Cesa_2017_Two-qubit} (see Fig.~\ref{fig:neighboring}). This approach is widely used, thanks to its straightforward implementation on existing platforms. However, its scalability is severely limited by the huge overhead in the gate count, leading to an exponential degradation of fidelity with the number of qubits. Indeed, the average number of physical gates required to realize a single logic gate scales linearly with the geometrical array size $L$, and the success probability of the logic gate reads
\begin{equation}\label{eq: fidelity nn}
\mathcal{F}_{\rm neighbor}\sim e^{-p_2\,L},
\end{equation}
where $p_2$ is the error probability of a physical two-qubit gate. While error correction can remedy this fault \cite{Fowler_2012_Proof}, the burden of the associated qubit and gate overheard can become practically prohibitive.

One way to overcome this {\it connectivity hurdle}  is to use transportable qubits~\cite{calarco2004quantum,kaufman2015entangling, Brickman_2009_Ultracold,Bluvstein_2022_Quantum,Tan_2022_Qubit,Bluvstein_2024_Logical,Bluvstein_2024_Logical,wang2024atomique,reichardt2024fault,Manetsch_2025_Tweezer,Radnaev_2025_Universal}.\footnote{There are other promising but less developed approaches involving optical couplers between qubits, see e.g.  \cite{Pellizzari_1995_Decoherence,Ramette_2022_Any-To-Any,sunami2025scalable,Shadmany_2025_Cavity,Sinclair_2025_Fault-tolerant,Grinkemeyer_2025_Error-detected}. They are not considered in the present paper.} 
Tweezer arrays appear to be a platform particularly convenient for implementing this approach thanks to their exceptional reconfiguration capabilities. In the most straightforward realization, a pair of transportable qubits can be moved close to each other whenever a two-qubit gate between these qubits should be performed. 
This idea is the basis for the reconfigurable Rydberg array architecture ~\cite{Bluvstein_2022_Quantum, Tan_2022_Qubit, Evered_2023_High-fidelity, Tan_2023_Compiling, Bluvstein_2024_Logical,wang2024atomique,reichardt2024fault,Radnaev_2025_Universal,bluvstein2025architectural,rines2025demonstration}. This approach is under active development, with massively parallel high-fidelity two-qubit gates \cite{Bluvstein_2022_Quantum}, and error correction codes in the fault-tolerant regime~\cite{Bluvstein_2024_Logical,reichardt2024fault,bluvstein2025architectural,rines2025demonstration} already demonstrated.

In the present work, we focus on a different way to use transportable qubits. Specifically, we explore architectures with two types of qubits -- {\it computational} and {\it messenger} \cite{Brickman_2009_Ultracold} -- with different roles. Computational qubits are essentially conventional qubits ready to be used to execute a quantum circuit. Importantly, they reside in a static tweezer array and do not move.  Disposable messenger qubits move between distant computational qubits and mediate effective two-qubit gates, thus enabling all-to-all connectivity with a size-independent overhead.

We propose five concrete architectures that realize this concept.
These architectures are described in Sec.~\ref{section II}.
In Sec.~\ref{section III}, we compare the messenger qubit approach to the reconfigurable array approach, discuss  advantages and challenges of each of the proposed architectures and analyze the technological advances required for their future development.
We conclude in Sec.~\ref{sec:conclusion}.



\section{Five architectures \label{section II}}

\begin{figure*}
\centering
\begin{minipage}{0.66\textwidth}
  \includegraphics[width=\linewidth]{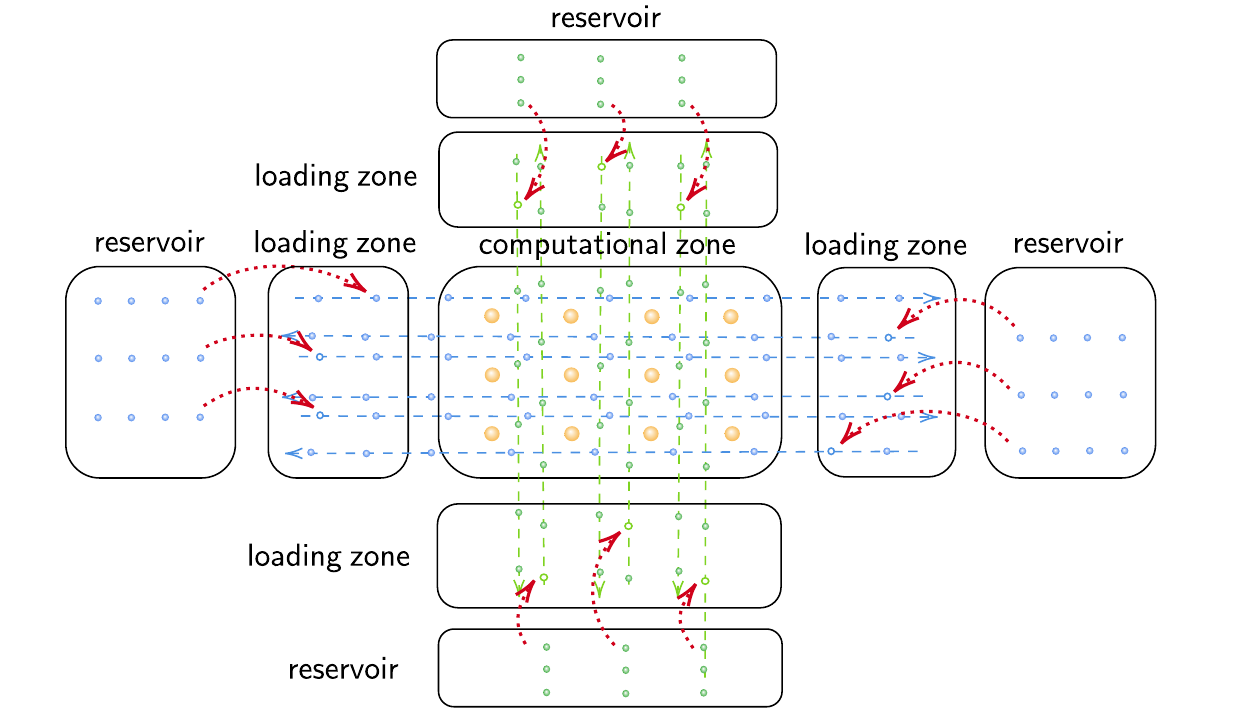}
\end{minipage}\\[2em]
\begin{minipage}{0.32\textwidth}
  \includegraphics[width=\linewidth]{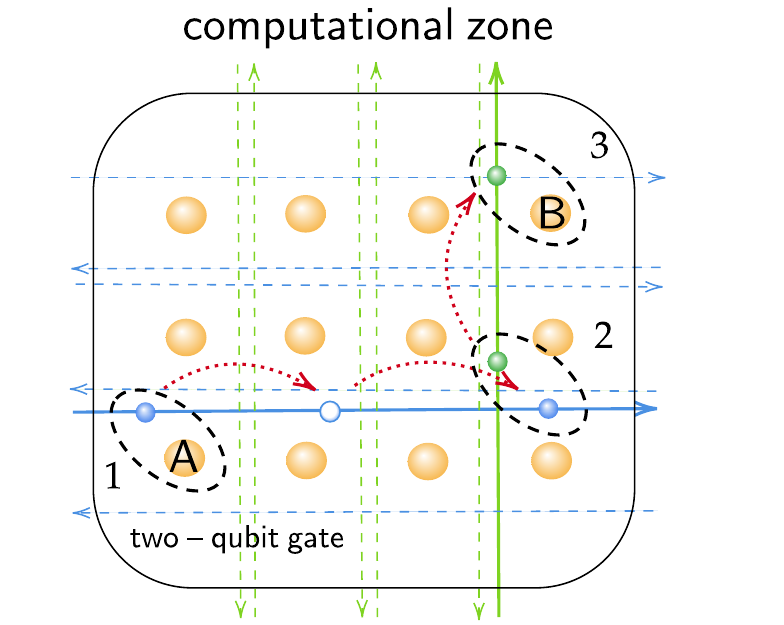}
\end{minipage}
\begin{minipage}{0.32\textwidth}
  \includegraphics[width=\linewidth]{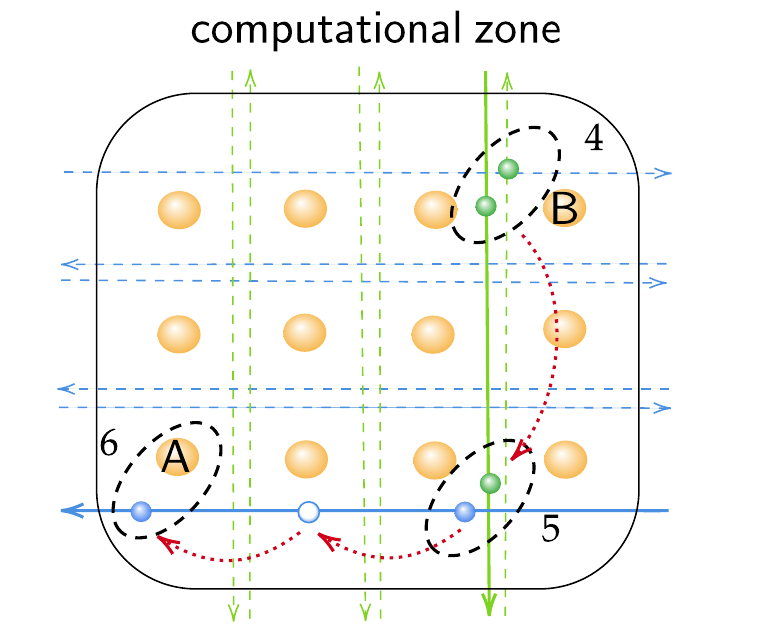}
\end{minipage}
\begin{minipage}{0.34\textwidth}
  \includegraphics[width=\linewidth]{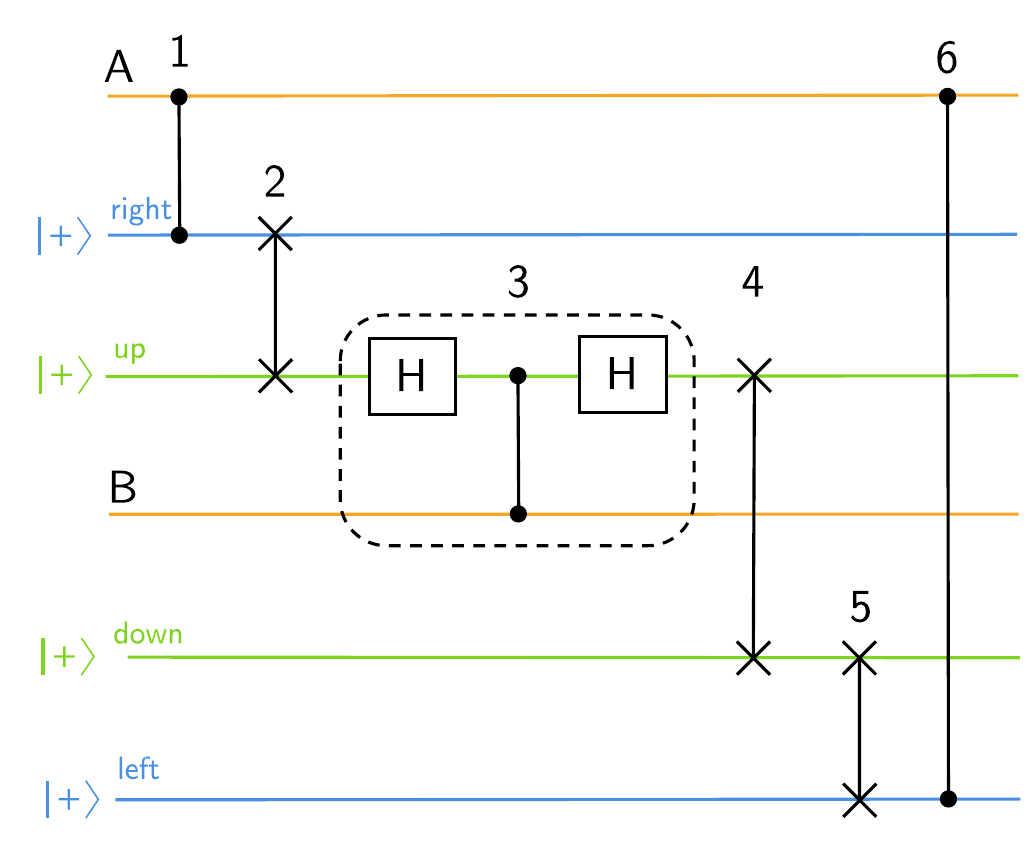}
\end{minipage}
\caption{Two-way conveyor belt architecture. {\bf Upper row:} general layout. Computational qubits (shown in orange) reside at rest in an immobile tweezer array in the computational zone. Messenger qubits (shown in blue or green,  depending on the  direction of movement) are shuttleged by atomic conveyor belts realized by moving tweezer arrays. There are four conveyor belts: two of them move in the opposite horizontal directions (shown in blue) and two -- in the opposite vertical directions. The messenger qubits are dynamically loaded to the conveyor belts from reservoirs in the loading zones. 
{\bf Lower row:} a sequence of physical nearest-neighbor two-qubit gates implementing a logical two-qubit gate between  target computational qubits,  A and B, located in the opposite corners of the array.
The physical  two-qubit gates are shown by dashed black ellipses and enumerated in the order of execution. In total, four messenger qubits from four different conveyor belts are employed to entangle two target computational qubits.}
\label{fig:two-way}
\end{figure*}


\subsection{Setting the stage}

We introduce two types of atomic qubits, which we refer to as {\it computational} and {\it messenger} \cite{Brickman_2009_Ultracold}. These types of qubits can be realized by a single atomic species, two different species~\cite{Shlyapnikov2015} or even multiple species.

Computational qubits are essentially conventional qubits residing in static tweezer arrays. 

Messenger qubits move between distant computational qubits and mediate effective two-qubit gates between the computational qubits. In general, we assume messenger qubits to be disposable, i.e. a given messenger qubit is used to mediate a single logical gate and is disposed afterwards. Technically, movable atoms can be reused, but only after re-initialization.

A major consequence of delegating interaction mediation to moving qubits is that the gate count overhead (the number of physical gates required to implement a logic gate) does not depend on the geometrical size of the processor and number of qubits. 
In this way, the connectivity hurdle can be resolved.

In general, optical tweezer arrays can have versatile one-, two- and three-dimensional geometries \cite{Browaeys2018}.
Furthermore, atoms or molecules residing in the array can serve as qubits or qudits \cite{Saffman2010}. 
To be specific, we focus on the most common two-dimensional layout, where computational  qubits reside in a square array of size $(a\,L)\times (a\,L)$ with $N=L^2$ qubits~\cite{Browaeys2016, Browaeys2016-2, Lukin2016}, where $a$ is the lattice spacing of the array. 

Apart from the {\it computational zone} containing the computational qubits, we introduce other zones serving various purposes --- e.g. {\it loading} or {\it measurement} zones. Partitioning a quantum processor into distinct zones has recently emerged as a prevalent strategy in neutral-atom quantum computing \cite{Bluvstein_2022_Quantum,reichardt2024fault,reichardt2024fault,Radnaev_2025_Universal,bluvstein2025architectural,rines2025demonstration}.

The key prerequisite for our approach is the availability of individually addressable single- and two-qubit gates. 
Recent theoretical and experimental advances give a clear promise that such gates can be realized with high fidelities and in parallel~\cite{Isenhower_2010_Demonstration,Jenkins_2022_Ytterbium,Graham_2022_Multi-qubit,Bluvstein_2022_Quantum,Lis_2023_Midcircuit,Sola_2023_Two-qubit,Radnaev_2025_Universal,bezuglov2024high}. 

To be specific, in each architecture we demonstrate a logical two-qubit \textsf{CZ} gate between two distant computational qubits located in the opposite corners of the square atomic array. Our approach works best when we assume that two types of {\it physical} two-qubit gates are accessible, the target logical gate (in our case, the \textsf{CZ} gate) and the \textsf{SWAP} gate.  The \textsf{CZ} gate has been implemented with errors well below  $10^{-2}$, with clear prospects for improvements \cite{Theis_2016_High-fidelity,Madjarov_2020_High,Radnaev_2025_Universal,Muniz_2025_High-Fidelity,Peper_2025_Spectroscopy,Finkelstein_2024_Universal,Tsai_2025_Benchmarking}.  While the \textsf{SWAP} gate is yet to be demonstrated experimentally, theoretical proposals \cite{Schuch_2003_Natural,Wu_2012_Quantum,Ni_2018_Dipolar,Wu_2022_Unselective,Xiao_2024_Effective,Wu_2024_Soft-controlled,Sun_2024_Holonomic,Li_2024_High-tolerance,Wang_2025_Design,Wang_2025_Deterministic} suggest that it can also be implemented with low errors. Therefore, in the main text we present results assuming that both \textsf{CZ} and \textsf{SWAP} gates are natively available at the hardware level. We also assume the availability of arbitrary single-qubit gates and measurements. More details on the required technological stack are presented in Sec. \ref{sec:stack}.

In the following, we describe five specific architectures presented in Figs. \ref{fig:two-way}--\ref{fig:catch-throw-mes}. 
They differ in the specific way of implementing logical two-qubit gates between the computational qubits with the help of moving messenger qubits. 
In Secs.~\ref{sec: two-way}-\ref{sec: throw-and-measure}, we introduce these architectures one by one. 
In Secs.~\ref{sec: fidelity} and \ref{sec: time}, we provide unified estimates for logical two-qubit gate fidelity and run time, respectively.




\subsection{Two-way conveyor belt architecture \label{sec: two-way}}

This architecture is illustrated in Fig.~\ref{fig:two-way}. 
The messenger qubits are moved by atomic {\it conveyor belts}~\cite{Schrader_2001_Optical,  Hickman_2020_Speed,Bluvstein_2022_Quantum}, which are moving tweezer arrays (or, alternatively, optical lattices) filled with atoms. 
In this architecture, there are four conveyor belts that move with constant velocities in opposite vertical and opposite horizontal directions. The conveyor belts overlay the fixed array hosting the computational qubits in such a way that a Rydberg gate can be performed between a computational qubit and a messenger qubit passing by, as well as between two messenger qubits moving in opposite or orthogonal directions. The messenger qubits are either preloaded into the moving arrays prior to computation or, preferably, continuously loaded from the {\it reservoirs} during computation in the {\it loading zone} (see, e.g., Ref.~\cite{Norcia_2024_Iterative}). Each messenger qubit is initialized in the $|+\rangle$ state. This initialization takes place in the loading zone and can be accomplished by applying a global Hadamard single-qubit gate to qubits in the state $\ket{0}$. Messenger qubits that have left the computational zone can be discarded or recycled in the reservoirs. 

The logic \textsf{CZ} gate between the two target computational qubits is implemented through a sequence of  $n_2=6$ nearest-neighbor two-qubit gates (3 \textsf{CZ} gates and 3 \textsf{SWAP} gates) and $n_1=2$ Hadamard single-qubit gates applied to  computational and messenger qubits,  as shown in Fig. \ref{fig:two-way}.



\subsection{One-way conveyor belt architecture}

\begin{figure*}
\begin{minipage}{0.66\textwidth}
\includegraphics[width=\linewidth]{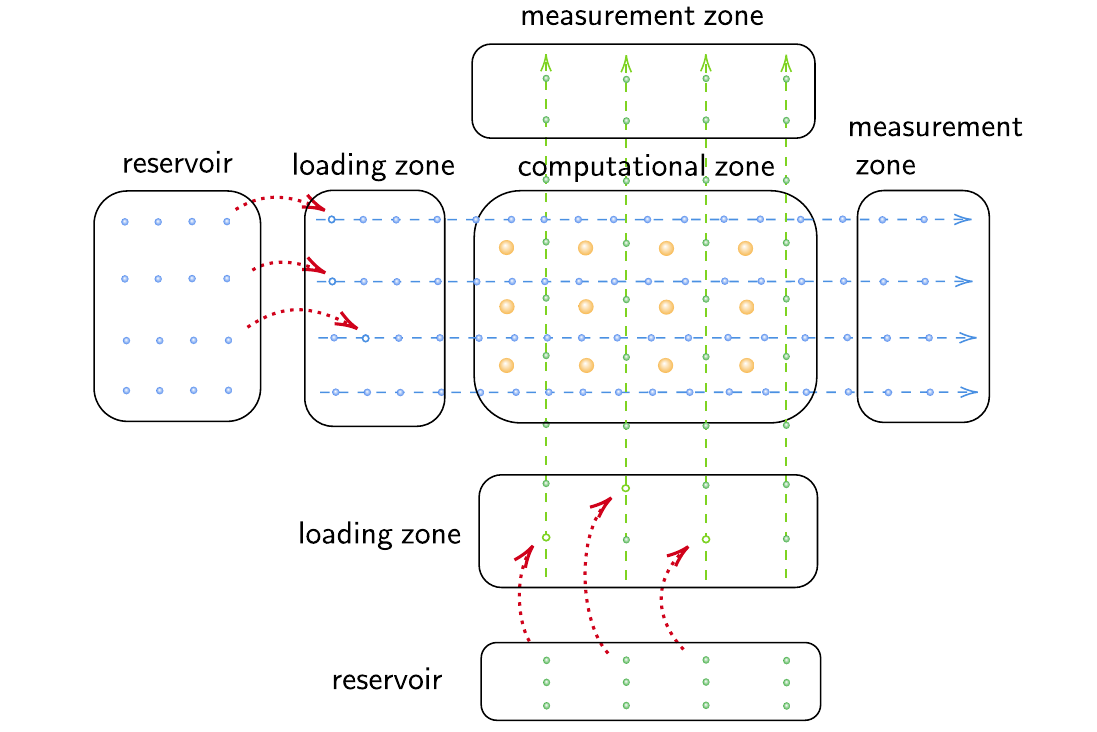}
\end{minipage}\\[2 em]
\begin{minipage}{0.33\textwidth}
  \includegraphics[width=\linewidth]{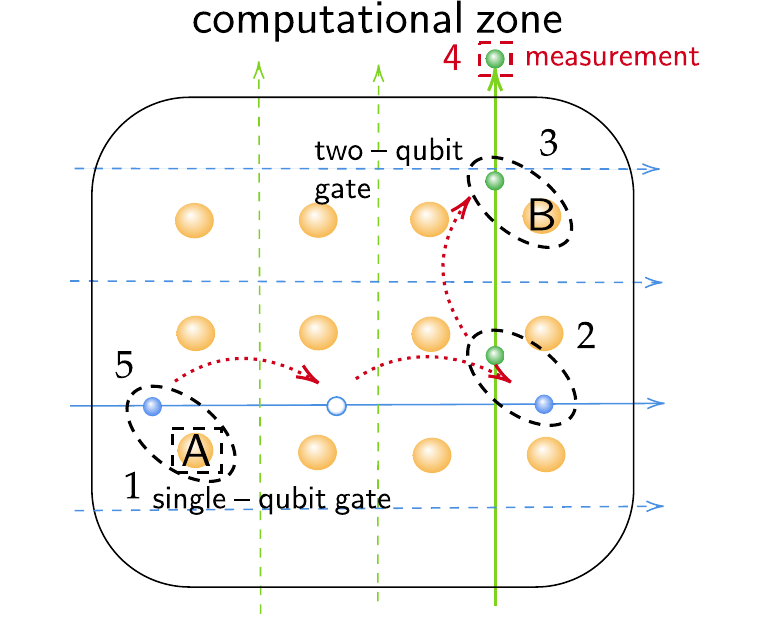}
\end{minipage}
~~
\begin{minipage}{0.32\textwidth}
  \includegraphics[width=\linewidth]{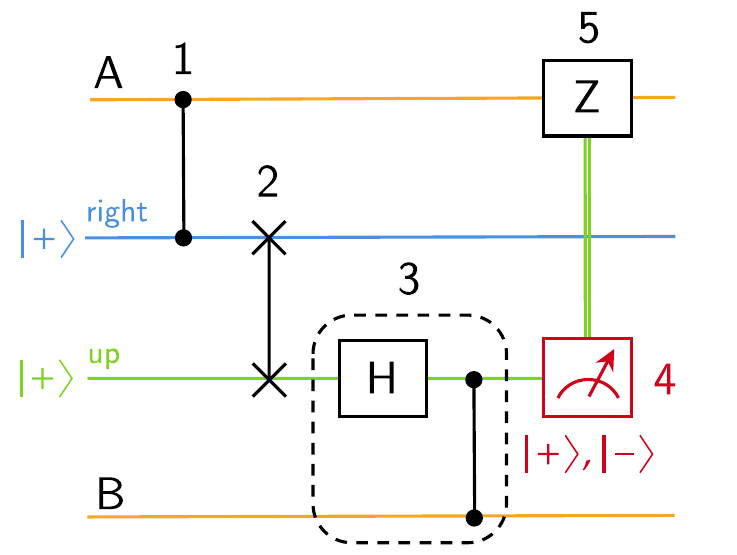}
\end{minipage}\\[1 em]
\begin{minipage}{0.33\textwidth}
  \includegraphics[width=\linewidth]{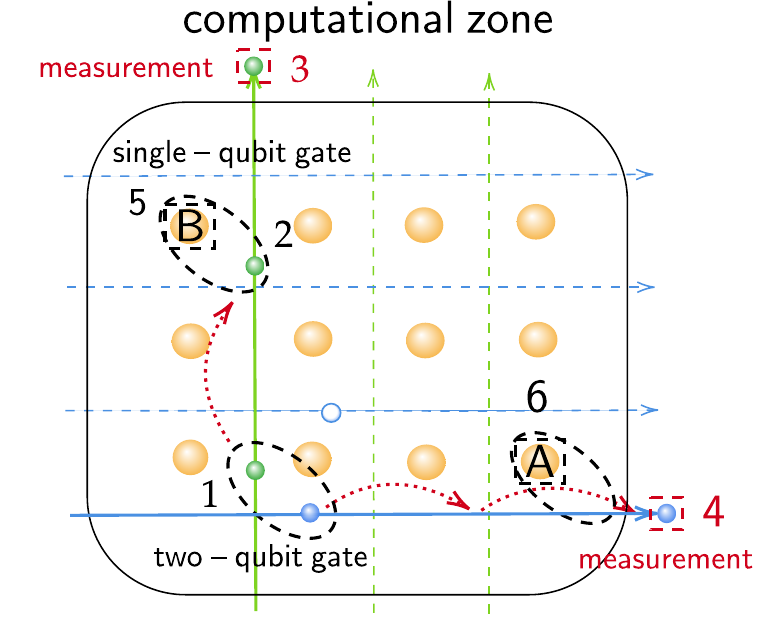}
\end{minipage}
~~
\begin{minipage}{0.32\textwidth}
  \includegraphics[width=\linewidth]{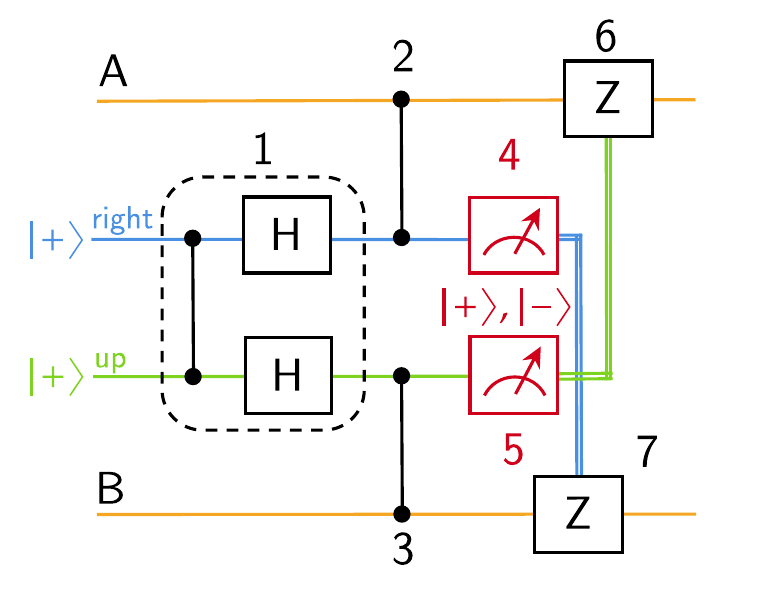}
\end{minipage}
\caption{One-way conveyor belt architecture.
{\bf Upper row:} general layout. Analogously to the two-way conveyor belt architecture shown in Fig.~\ref{fig:two-way}, messenger qubits are shuttleged by atomic conveyor belts moving with constant velocities. However, here there are only two conveyor belts (and two loading zones) instead of four. The reverse flow of information is accomplished by a quantum teleportation protocol involving measurements over  messenger qubits and conditional single-qubit gates over computational qubits. Measurements are performed in two  readout zones.     
{\bf Middle and lower row:}  two sequences of physical gates and measurements implementing a logical two-qubit gate between  target computational qubits for two nonequivalent relative locations of target qubits.
The physical  two-qubit gates, single-qubit gates and single-qubit measurements are shown by dashed black ellipses, black squares and  red squares, respectively. The measurements are performed in  the basis $\ket{\pm} = \left(\ket{0}\pm\ket{1}\right)/\sqrt{2}$.}
\label{fig:one-way}
\end{figure*}

This architecture is shown in Fig.~\ref{fig:one-way}. 
In contrast to the previous case, there are only two square lattice conveyor belts that move in orthogonal directions 
(e.g., one from left to right and another from bottom to top). 
A two-qubit gate between the target qubits is communicated by means of a quantum teleportation protocol~\cite{Bennett_1993_Teleporting} involving a messenger qubit measurement and a subsequent conditional single-qubit gate over a logical qubit. In fact, the precise physical gate sequence depends on the  relative location of the target qubits, as shown in Fig.~\ref{fig:one-way}. 
In any case, there are  $n_2=3$ two-qubit physical gates per logic gate. 
In addition, there are either $n_{\rm r}=1$ or $n_{\rm r}=2$ single-qubit measurements and, respectively, at most $n_{1}=2$ or $n_{1}=4$ single-qubit gates, depending on the relative qubit location, see Fig.~\ref{fig:one-way}. The processor layout includes two loading zones and two  readout zones.

Compared to the previous architecture, where three two-qubit gates are used to ``return'' the entanglement from the target qubit $B$ to the qubit $A$, here these gates are replaced by the measurement and the conditional single-qubit gate. This trade-off can be beneficial when a fast and reliable qubit readout is available~\cite{Bergschneider_2018_Spin-resolved,Xu_2021_Fast,Deist_2022_Mid-Circuit,Grinkemeyer_2025_Error-detected}, which is discussed in more detail in Sec. \ref{sec: time}. 
An additional benefit of this architecture is the reduced complexity of the moving tweezer array arrangement and, as a potential consequence, the reduced cross-talk while performing two-qubit gates.

\begin{figure*}[t]
\begin{minipage}{0.66\textwidth}
\includegraphics[width=\linewidth]{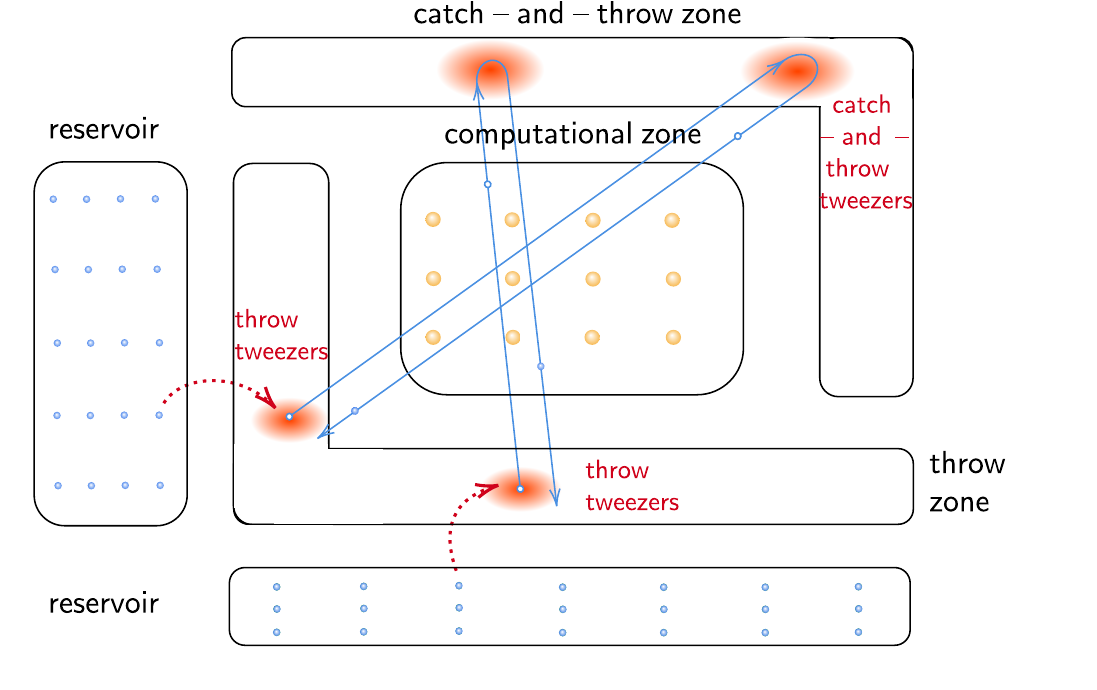}
\end{minipage}\\[2 em]
\begin{minipage}{0.33\textwidth}
  \includegraphics[width=\linewidth]{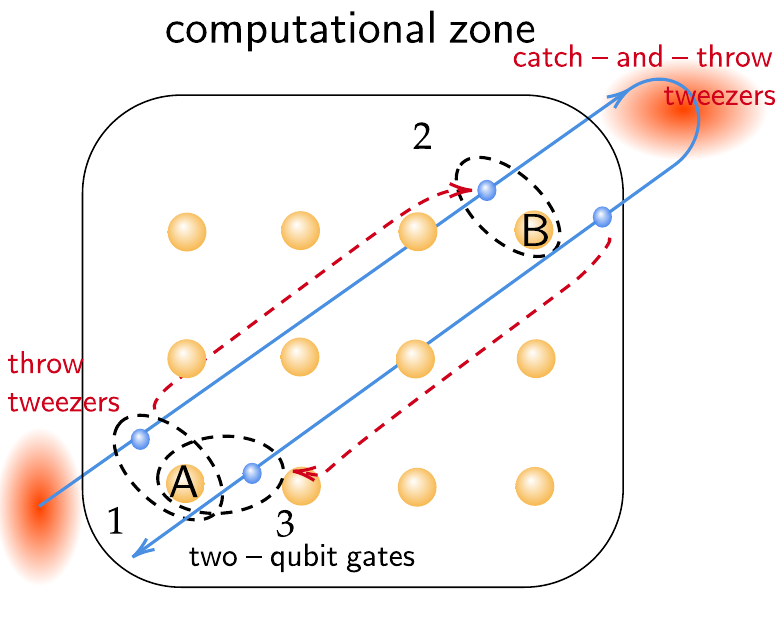}
\end{minipage}
\begin{minipage}{0.32\textwidth}
  \includegraphics[width=\linewidth]{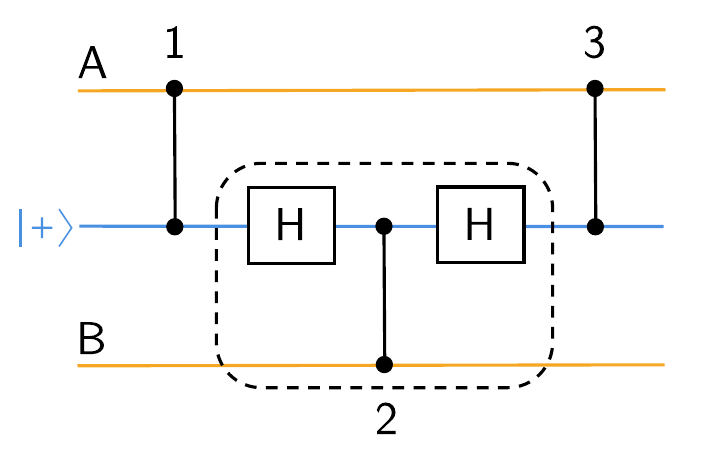}
\end{minipage} 
\caption{Throw-catch-throw architecture.  {\bf Upper row:} general layout. A messenger qubit is launched by purpose-built optical tweezers operating  within the throw zone and flies freely through the computational zone. Its trajectory passes by two target computational qubits, A and B. Two two-qubit gates are performed between the messenger and each of the target computational qubits during this passage. Then the messenger qubit enters the catch-and-throw zone, where it is decelerated (``caught'') and re-launched towards the target computational qubit A by catch-and-through tweezers~\cite{Hwang_2023_Optical}. The third two-qubit gate is performed between the messenger qubit and computational qubit A during the backwards passage.   {\bf Lower row:}  A sequence of physical gates  implementing a logical two-qubit gate between  target computational qubits. 
\label{fig:throw-catch-throw}
}
\end{figure*}

\begin{figure*}[t!]
\begin{minipage}{0.5\textwidth}
\includegraphics[width=\linewidth]{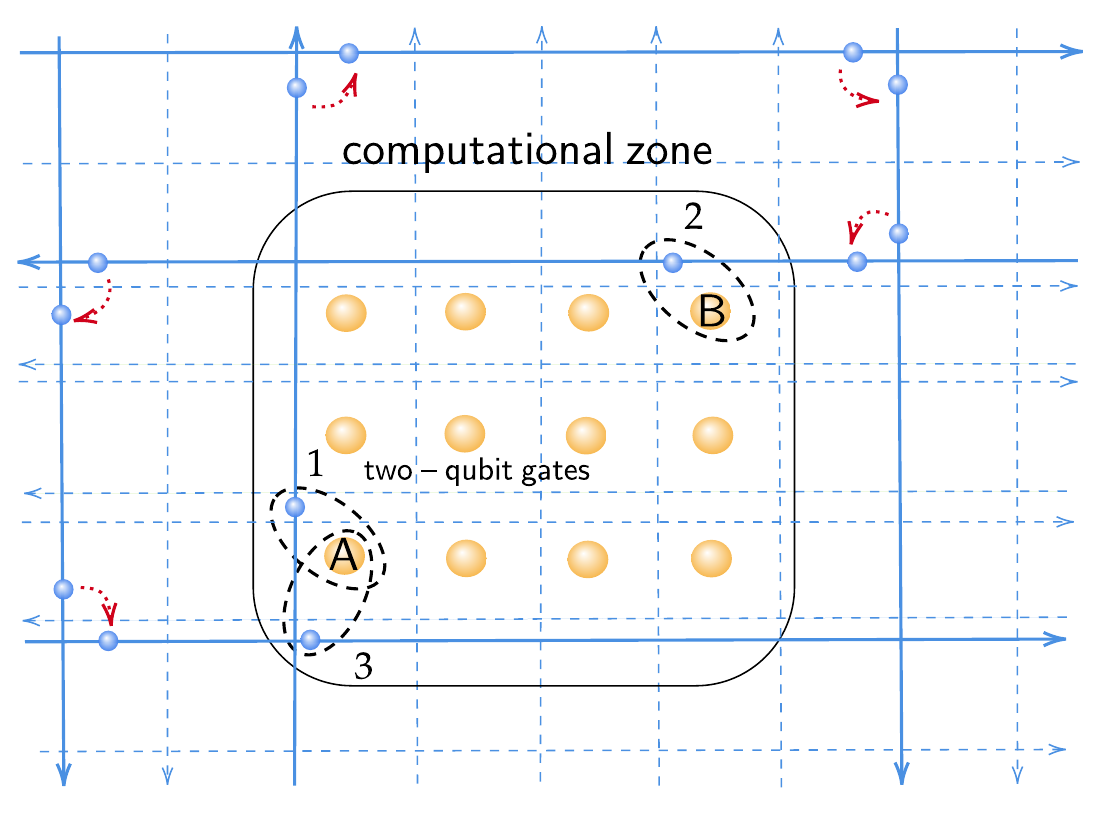}
\end{minipage}\\[2 em]
\caption{Shuttle-and-route architecture. A single messenger qubit is used to implement a logical qubit gate between computational qubits A and B. This messenger qubit is shuttled by conveyor belts through the computational zone and transferred from one conveyor belt to another, orthogonal one, outside the computational zone. A sequence of physical gates  implementing a logical two-qubit gate between  target computational qubits is identical to that for the throw-catch-throw architecture, {\it cf.} Fig. \ref{fig:throw-catch-throw}.
}
\label{fig:shuttle-and-route}
\end{figure*}

\begin{figure*}[t!]
\begin{minipage}{0.66\textwidth}
\includegraphics[width=\linewidth]{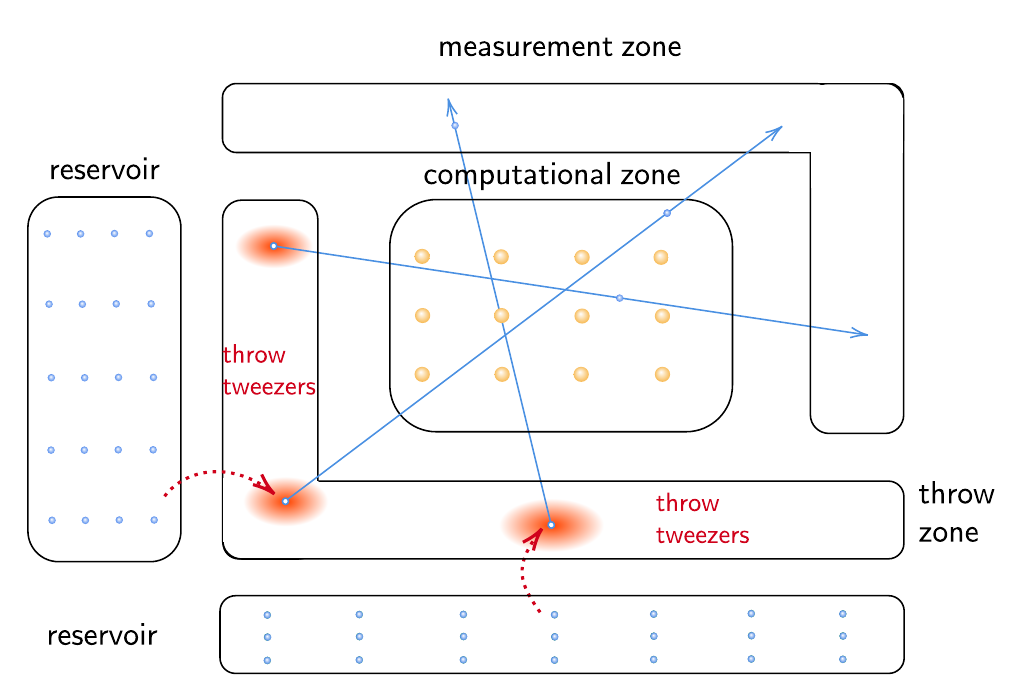}
\end{minipage}\\[2 em]
\begin{minipage}{0.32\textwidth}
  \includegraphics[width=\linewidth]{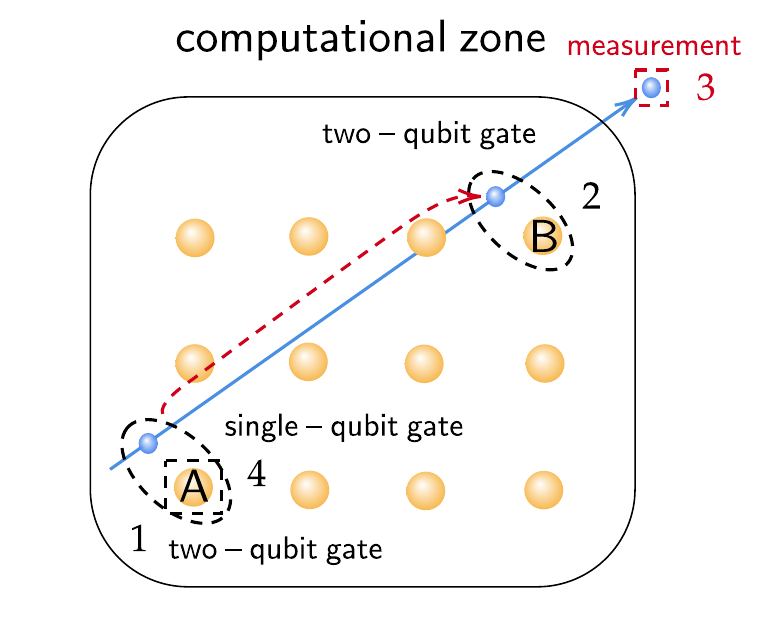}
\end{minipage}
\begin{minipage}{0.34\textwidth}
  \includegraphics[width=\linewidth]{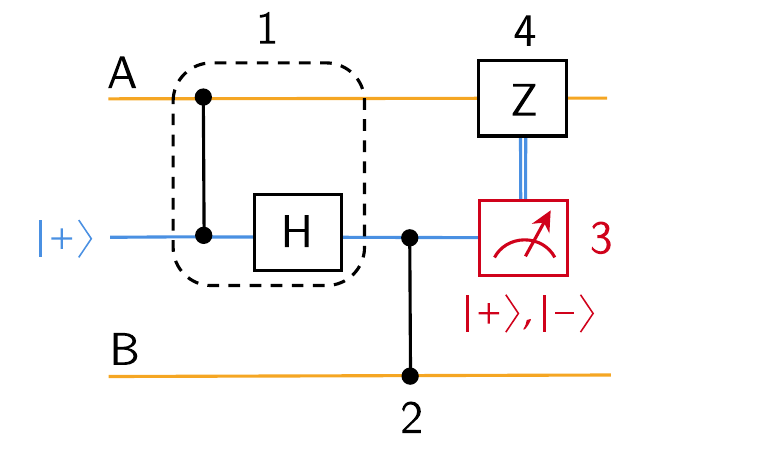}
\end{minipage} 
\caption{Throw and measure architecture.  {\bf Upper row:} general layout. Analogously to the throw-catch-throw architecture in Fig.~\ref{fig:throw-catch-throw}, a messenger qubit is launched by optical tweezers in the throw zone and flies freely through the computational zone, passing by two target computational qubits, A and B, with two-qubit gates performed upon the passage. After leaving the computational zone, the messenger qubit enters the readout zone where it is measured in the basis $\ket{\pm} = \left(\ket{0}\pm\ket{1}\right)/\sqrt{2}$, analogously to the one-way conveyor belt architecture in Fig.~\ref{fig:one-way}. Then  a conditional single qubit gate  is performed upon the qubit A.  {\bf Lower row:}  A sequence of physical gates  implementing a logical two-qubit gate between  target computational qubits.}
\label{fig:catch-throw-mes}
\end{figure*}



\subsection{Throw-catch-throw architecture}

Here, messenger qubits are not shuttleged by optical tweezers through the computational zone but instead freely fly through it, being launched, decelerated and relaunched by purpose-built  tweezers in the separate {\it throw} and {\it catch-and-throw } zones. In contrast to the previous architectures, here a single messenger qubit is used to mediate a two-qubit gate between distant computational qubits. This proceeds as follows. First, the messenger qubit is accelerated in the catch-and-through zone by movable optical tweezers and directed (``thrown'') along a trajectory passing near two target computational qubits. This enables two-qubit gates between the messenger qubit and each of the target computational qubit.  After that, the direction of the messenger qubit is inverted by the second tweezers in the catch-and-through zone, and the third two-qubit gate is performed when it passes the first target qubit for the second time. This gives $n_2=3$ physical two-qubit gates and $n_1=2$ single-qubit gates for one logic gate. As in the previous cases, the layout features a reservoir where the messenger qubits are sourced.  

A major component of this architecture are tweezers capable of throwing and catching messenger qubits. This component has recently been experimentally demonstrated~\cite{Hwang_2023_Optical}.  The state-of-the-art fidelity of not losing the messenger qubit in the throw-catch-throw process is $0.94$, with a clear potential for improvement~\cite{Hwang_2023_Optical}.

A scalable processor should have multiple throw and catch-and-throw tweezers to enable multiple two-qubit gates are implemented in parallel.

The advantage of the architecture is the absence of moving tweezers in the main zone, low cross-talk, and low two-qubit gate count. The main challenge of the architecture is the necessity for multiple throw and throw-and-catch optical tweezers capable of precisely directing messenger qubits.


\subsection{Shuttle-and-route architecture}

This architecture is illustrated in Fig. \ref{fig:shuttle-and-route}. 
It is a hybrid of the conveyor belt architecture and the throw-catch-throw architecture, combining some important advantages of both. As in the the throw-catch-throw architecture, only one messenger qubit is used per logical two-qubit gate, thus minimizing the two-qubit gate count. In fact, at the circuit level,  the decomposition of the logical two-qubit gate in physical gates is the same as in the throw-catch-throw architecture, {\it cf.} Figs. \ref{fig:throw-catch-throw} and \ref{fig:shuttle-and-route}. 

However, in contrast to the throw-catch-throw architecture, here this single messenger qubit is shuttleged by optical conveyor belts, analogously to the conveyor belt architectures.  This way one circumvents the need for catch-and-throw tweezers precisely directing messenger qubits to freely fly along chosen trajectories --- a technology not yet demonstrated at scale.

An important new element in this architecture is the {\it routing} operation — a process of switching the messenger qubit from one conveyor belt to another. This operation is performed at the intersection of two orthogonal conveyor belts and transfers the messenger qubit onto the new belt, thereby changing its direction of motion by 90 degrees. In general, five such routing operations are required per logical two-qubit gate, as illustrated in Fig.~\ref{fig:shuttle-and-route}.

Transferring atoms between  static and moving tweezer arrays has been demonstrated with an error around $10^{-3}$ per transfer \cite{Manetsch_2025_Tweezer}. We expect that that transferring atoms between two moving tweezer arrays can be performed with a similarly low error.


\subsection{Throw-and-measure architecture \label{sec: throw-and-measure}}

This architecture, illustrated in Fig.~\ref{fig:catch-throw-mes}, utilizes flying qubits as in the throw-catch-throw scheme and quantum teleportation of the messenger qubit state as in the one-way conveyor belt scheme. The messenger qubits are ejected by optical tweezers in the {\it throw zone} and follow a trajectory connecting two target computational qubits, with two-qubit gates performed on a passage. After the messenger qubit leaves the computational zone, it is measured in the readout zone, and the entanglement is teleported back to the first target qubit by means of a conditional single-qubit gate.

This architecture requires $n_2=2$ physical two-qubit gates, $n_{\rm r}=1$ measurement, and at most $n_1=2$ single-qubit gates (one gate being conditioned on the measurement outcome) per  logical two-qubit gate. In comparison to the throw-catch-throw architecture, here the catching step is traded for the measurement and the conditional single-qubit gate. 
As in the case of the one-way conveyor belt architecture, the implementation of the throw-and-measure architecture requires fast single-qubit measurements with high fidelity.



\subsection{Gate fidelity \label{sec: fidelity}}

For any of the above architectures, the fidelity $ \mathcal{F}$ of a logic two-qubit gate between two distant computational qubits can be represented as
\begin{equation}
    \mathcal{F} = 
               \mathcal{F}_2^{\,n_2}\times\mathcal{F}_1^{\,n_1}\times
\mathcal{F}_{\text{r}}^{\, n_{\mathrm r}}\times\mathcal{F}_{\text{shuttle}}.
      \label{eq:fidelity}
\end{equation}
Here, $\mathcal{F}_2$, $\mathcal{F}_1$ and $\mathcal{F}_{\text{r}}$ are, respectively, fidelities of physical  two-qubit gate, single qubit gate and single qubit readout, and  $\mathcal{F}_{\text{shuttle}}$ is the fidelity of shuttling all of the messenger qubits involved in the logical two-qubit gate without loosing them and altering their quantum states. Recall that $n_2$, $n_1$, and $n_{\mathrm r}$ are, respectively,  numbers of physical two-qubit and single-qubit gates and single-qubit readout per one logical two-qubit gate. These numbers for different architectures are summarized in Table~\ref{tab:Ngates}. Note that, in architectures involving messenger qubit measurement, some single-qubit gates are conditioned on the measurement result. In these cases $n_1$ is the maximal number of single-qubit gates, i.e. it refers to the worst case scenario.

The first three contributions in Eq.~\eqref{eq:fidelity} are manifestly independent of the geometrical size of the processor and the total number of qubits. 

The scaling of the shuttling error may be more subtle, as it contains several contributions—some manifestly independent of the system size and others potentially size-dependent. The size-independent contributions include errors associated with the trapping, initialization, and launching of messenger qubits. A size-dependent contribution can, in principle, arise from terms proportional to the time the messenger qubit spends either in the conveyor belt or in free flight, the latter being linear in the geometrical size of the processor. The physical origins of these terms may include heating by trapping laser light in the case of conveyor belts, and wave-packet spreading in the case of free flight. Nevertheless, we expect that, in the foreseeable future, the shuttling error will constitute the smallest contribution to the total error budget. This expectation is supported by state-of-the-art experimental results, which already demonstrate cumulative shuttling errors  around $10^{-3}$ per dozens of shuttling cycles~\cite{Manetsch_2025_Tweezer}. The shuttling fidelity will further  benefit from future improvements in   optimal shuttling~\cite{qi2021fast,Lam_2021_Demonstration, pagano2024optimal,hwang2024fast} and cooling~\cite{wang2024atomique} techniques.

It should be emphasized that in our proposal conveyor belts move with constant velocities, therefore the heating due to nonuniform center-of-mass motion occurs only at the loading  (and routing, for the shuttle-and-route architecture) step.

\begin{figure*}[t!]
\begin{minipage}{0.3\textwidth}
\includegraphics[width=\linewidth]{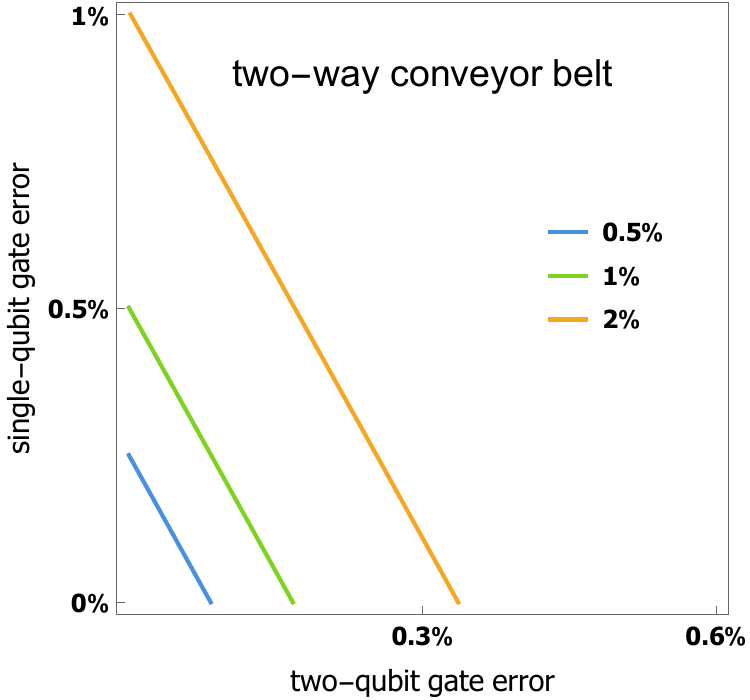}
\end{minipage}
\begin{minipage}{0.3\textwidth}
\includegraphics[width=\linewidth]{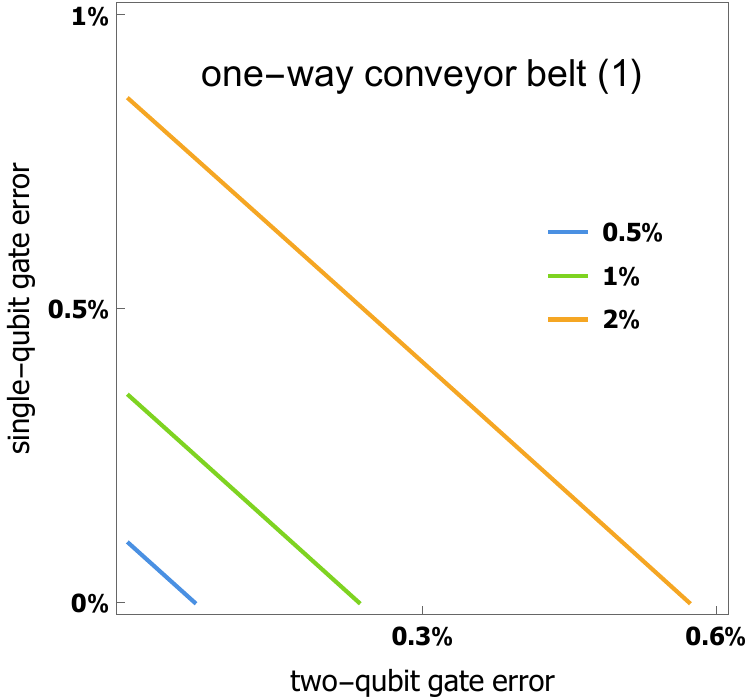}
\end{minipage}
\begin{minipage}{0.3\textwidth}
\includegraphics[width=\linewidth]{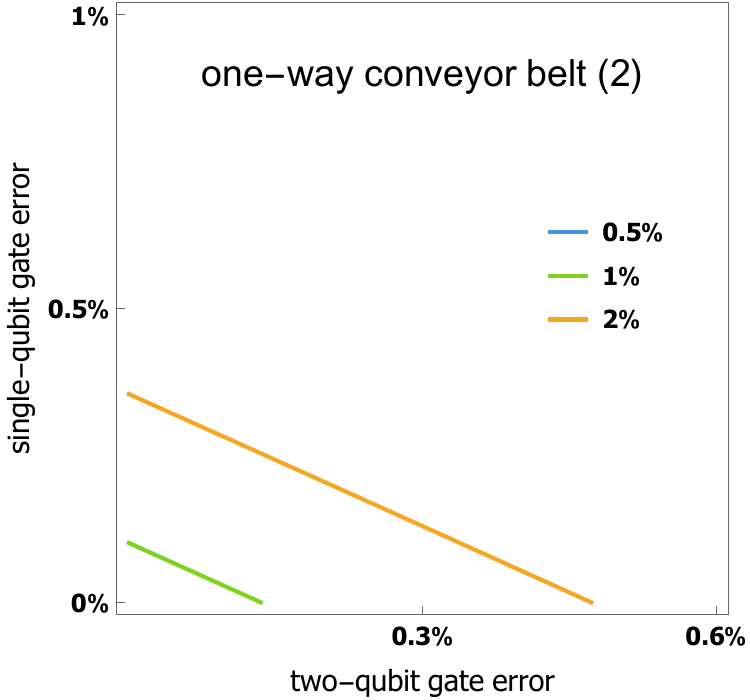}
\end{minipage}
\\[2 em]
\begin{minipage}{0.3\textwidth}
\includegraphics[width=\linewidth]{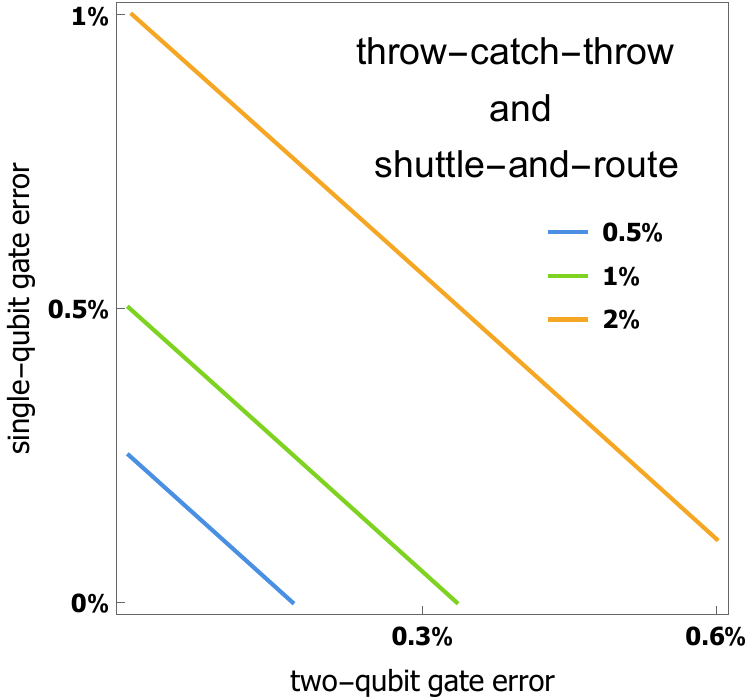}
\end{minipage}
\begin{minipage}{0.3\textwidth}
\includegraphics[width=\linewidth]{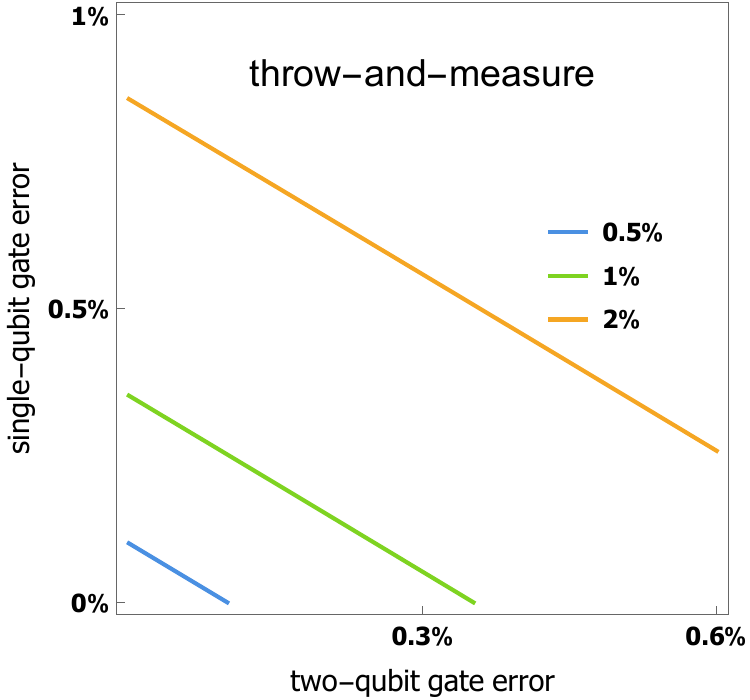}
\end{minipage}
\caption{Contribution of single- and two-qubit physical gate errors to the fidelity of a logical two-qubit gate between two distant computational qubits. The logical gate error is calculated for various architectures using Eq. (\ref{eq:fidelity}) and gate and readout counts in Table~\ref{tab:Ngates}. For the one-way conveyor belt architecture, two different plots correspond to the two different relative locations of target computational qubits, see Fig. \ref{fig:one-way}. The  qubit readout error is set to $3\times10^{-3}$. The errors of physical \textsf{CZ} and \textsf{SWAP} gates are assumed to be equal. 
 The shuttling and routing fidelities are set to unity. Under the latter assumption, the error budgets for the throw-catch-throw and shuttle-and-route architectures coincide.}
\label{fig:errors}
\end{figure*}

\begin{figure*}[t!]
\begin{minipage}{0.3\textwidth}
\includegraphics[width=\linewidth]{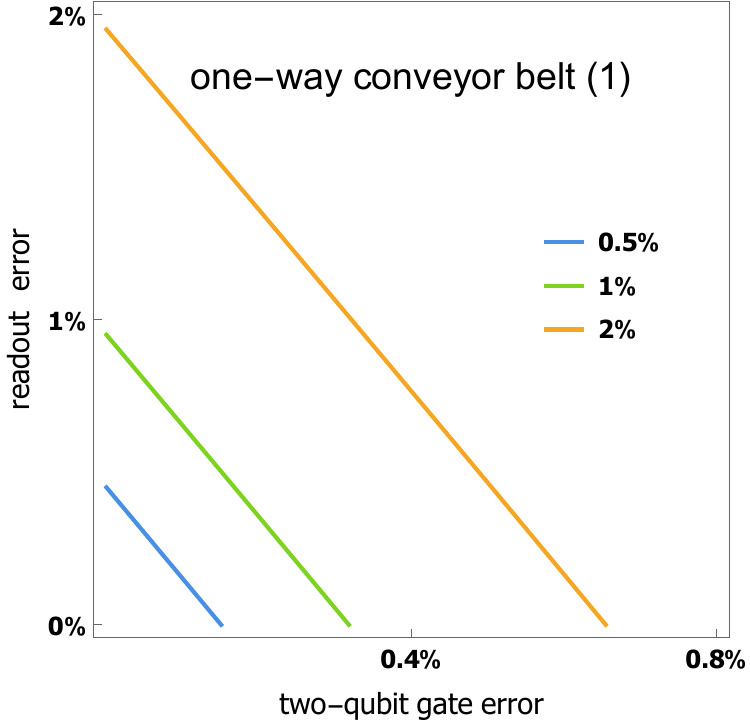}
\end{minipage}
\begin{minipage}{0.3\textwidth}
\includegraphics[width=\linewidth]{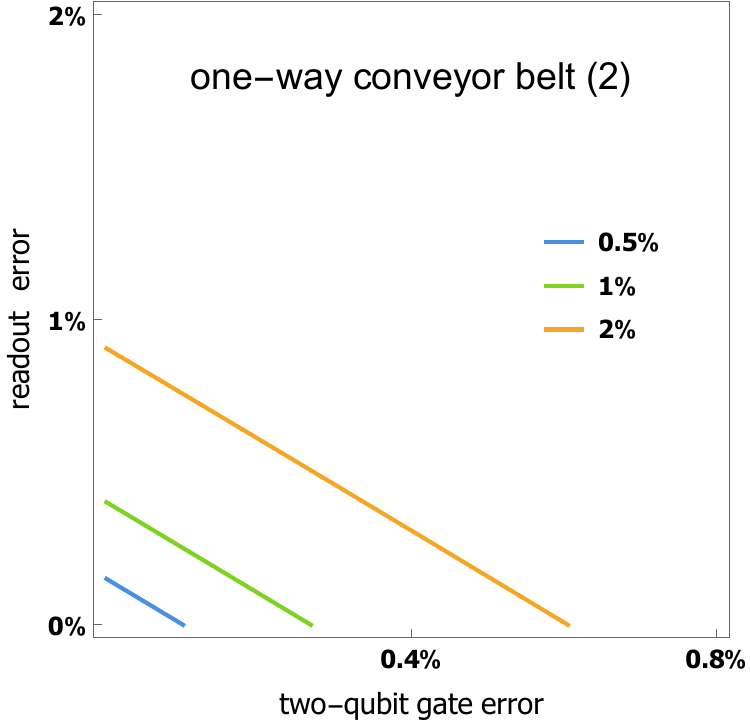}
\end{minipage}
\begin{minipage}{0.3\textwidth}
\includegraphics[width=\linewidth]{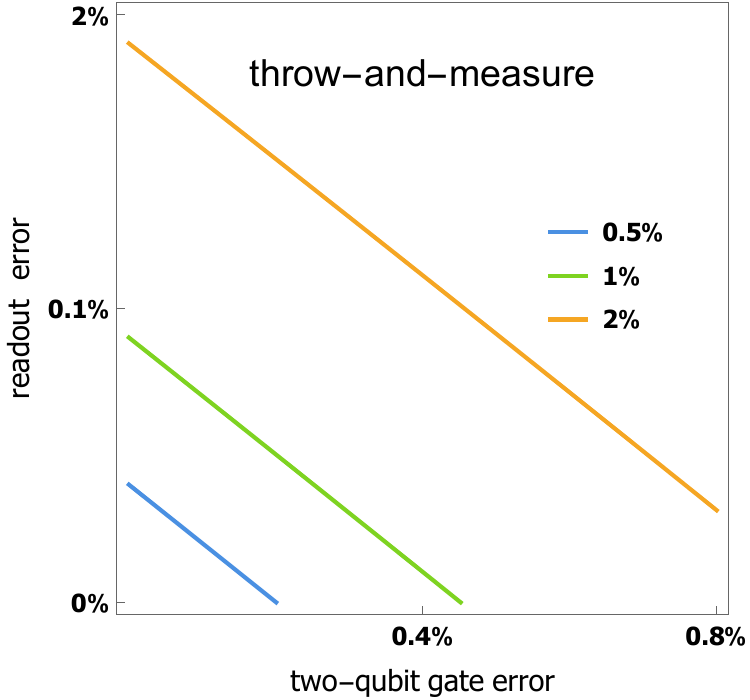}
\end{minipage}
\caption{Contribution of readout and two-qubit physical gate errors to the fidelity of a logical two-qubit gate between two distant computational qubits.  The logical gate error is calculated for various architectures using Eq. (\ref{eq:fidelity}) and  gate and readout counts in Table~\ref{tab:Ngates}. For the one-way conveyor belt architecture, two different plots correspond to the two different relative locations of target computational qubits, see Fig. \ref{fig:one-way}. The single-qubit error is set to $5\times10^{-4}$. The errors of physical \textsf{CZ} and \textsf{SWAP} gates are assumed to be equal. 
 The shuttling fidelity is set to unity.}
\label{fig:errors_measurement}
\end{figure*}

Figs.~\ref{fig:errors} and~\ref{fig:errors_measurement} illustrate the magnitudes of errors from various sources that are required to implement a logical two-qubit gate with an overall error around $10^{-2}$ -- a common near-term target for fault-tolerant quantum computing \cite{Wang_2011_Surface,Wu_2022_Erasure,Bluvstein_2024_Logical,Bravyi_2024_High-threshold,Katabarwa_2024_Early}. Evidently, the five proposed architectures differ substantially in their error requirements. Further discussion of various contributions to the error budget is presented in Sec.~\ref{sec:stack}.

\subsection{Gate time \label{sec: time}}

For architectures that do not involve the messenger qubit readout, the logical two-qubit gate time $t$ is merely the total shuttling time $t_{\rm shuttle}$ of the messenger qubit (qubits) implementing the logical gate. The shuttling time is limited by the messenger qubit velocity $v$ and can be estimated as $t_{\rm shuttle}\sim L\,a/v$.  
Importantly, there is no separate contribution of the physical two-qubit gate time, $t_2$, to the logical gate time. This is because two-qubit gates are executed on the fly, while the messenger qubit passes by the computational qubit. Instead, $t_2$ enters into $t_{\rm shuttle}$. In fact, the requirement to perform a physical two-qubit gate on the fly constitutes the major limitation on  the velocity of the messenger qubit. This velocity must be sufficiently low for the messenger qubit to remain within the Rydberg blockade radius, $R$, of the target computational qubit during the two-qubit interaction. Considering that, to avoid crosstalk, one typically requires $R$ to be smaller than the lattice spacing, $a$,\footnote{Relaxing this requirement would help reduce the shuttling time. This could potentially be achieved by employing two different atomic species for the computational qubits organized in the checkerboard pattern~\cite{Beterov_2015_Rydberg,Singh_2022_Dual-Element}.} one obtains the bound $v \lesssim a/t_2$ and an estimate

\begin{equation}\label{shuttle_time}
t=t_{\rm shuttle} \sim L\,t_2.
\end{equation}

Note that this estimate coincides with the estimated logical two-qubit gate time for implementations based on sequences of nearest-neighbor two-qubit gates. \cite{Arute_201_Quantum,Jeong_2022_Rydberg,sun2024buffer, he2024distant,Cesa_2017_Two-qubit,giudici2025fast,PhysRevX.15.011009}. 

Typical state-of-the-art values for physical two-qubit gate times are $t_2 \sim 1~\mu\mathrm{s}$~\cite{Madjarov_2020_High,Bluvstein_2022_Quantum,reichardt2024fault,schmid2024computational,Wintersperger_2023_Neutral,Radnaev_2025_Universal}. Combined with a typical lattice spacing, $a$, of a few micrometers, this implies achievable shuttling velocities on the order of meters per second. Comparable velocities have been demonstrated in state-of-the-art experiments with reconfigurable atom arrays \cite{Bluvstein_2022_Quantum,Bluvstein_2024_Logical,reichardt2024fault,Radnaev_2025_Universal,Manetsch_2025_Tweezer}.

In the catch-and-throw architecture, $t_{\rm shuttle}$ acquires an additional contribution from the catch-and-throw step. Similarly, in the shuttle-and-route architecture, routing operations add to the shuttling time. These contributions are independent of the system size and are therefore not included in the above rough estimate. Based on current experimental experience~\cite{Bluvstein_2022_Quantum,Bluvstein_2024_Logical,reichardt2024fault,Radnaev_2025_Universal,Manetsch_2025_Tweezer}, they are expected to remain below a few microseconds.

Finally, in the one-way conveyor belt and throw-and measure architectures, there is another contribution to logical two-qubit gate: the time required to readout the messenger qubit and perform a subsequent conditional single-qubit rotations. Conventional atomic qubit readout typically takes hundreds of microseconds \cite{schmid2024computational,Wintersperger_2023_Neutral}, which can be too long for practical implementations. Therefore these architectures seem viable provided one utilizes unconventional schemes of fast qubit readout, where readout times are orders of magnitude lower \cite{Bochmann_2010_Lossless,volz2011measurement,Xu_2021_Fast,Deist_2022_Mid-Circuit,Petrosyan_2024_Fast,Grinkemeyer_2025_Error-detected}, see Sec. \ref{sec:stack} for further discussion.

\begin{table}[t!]
  \begin{center}
    \begin{tabular*}{0.45\textwidth}{@{\extracolsep{\fill}}l|c|c|c} 
      \textbf{architecture} & $n_1$ & $n_2$ & $n_{\mathrm r}$\\
      \hline 
      two-way conveyor belt  & 2 & 6 & 0\\[0.5 em]
      one-way conveyor belt (1) & 2 & 3 & 1\\[0.2 em]
     one-way conveyor belt (2) & 4 & 3 & 2\\[0.5 em]
      throw-catch-throw  & 2 & 3 & 0\\[0.5 em]
      shuttle-and-route  & 2 & 3 & 0\\[0.5 em]
      throw-and-measure   & 2 & 2& 1\\
    \end{tabular*}
  \end{center}
  \caption{\label{tab:Ngates}The number of single-qubit physical gates, $n_1$, two-qubit physical gates, $n_2$, and single-qubit readout, $n_{\mathrm r}$, required to perform  a logical two-qubit gate in different architectures. For the  one-way conveyor belt, the gate counts vary for two inequivalent relative locations of target qubits, as shown in Fig. \ref{fig:one-way} }
\end{table}


\section{Discussion \label{section III}}

\subsection{Comparison to the reconfigurable Rydberg array architecture}

In recent years, the reconfigurable Rydberg array architecture has shown remarkable progress and clear scalability in qubit count~\cite{Bluvstein_2022_Quantum, Evered_2023_High-fidelity,Bluvstein_2024_Logical,reichardt2024fault,Radnaev_2025_Universal,bluvstein2025architectural,rines2025demonstration}, with fault-tolerant operations already demonstrated on systems comprising hundreds of qubits~\cite{Bluvstein_2024_Logical,reichardt2024fault,bluvstein2025architectural,rines2025demonstration}.

However,  scalability in another dimension   --- circuit depth ---  remains challenging. Compared with static layouts, architectures based on atomic rearrangement introduce an additional contribution to the error budget that may limit scalability with increasing circuit depth. Each two-qubit gate requires a sequence of atomic moves in which a qubit is accelerated and decelerated, leading to heating of its motional degrees of freedom. While negligible for shallow circuits~\cite{Bluvstein_2022_Quantum}, this heating error accumulates as circuits deepen~\cite{reichardt2024fault,Manetsch_2025_Tweezer}, eventually degrading gate fidelities below the fault-tolerance threshold~\cite{Robicheaux_2021_Photon-recoil,Manetsch_2025_Tweezer}. Importantly, there exists a fundamental trade-off between the heating rate and the gate time, constraining the extent to which improved transport protocols can mitigate this error~\cite{Lam_2021_Demonstration}. Addressing this limitation will likely require the ability to cool atoms during computation without disrupting quantum coherence, as proposed in Refs.~\cite{Reichenbach_2007_Sideband,Belyansky_2019_Nondestructive,Graham_2023_Midcircuit}, or, perhaps, the development of even more sophisticated error-mitigation techniques.

In contrast, in our approach the computational qubits remain stationary and are thus unaffected by motional heating. For the messenger qubits, heating occurs only during acceleration or rerouting but does not accumulate with circuit depth, since the messenger qubits are discarded after each logical two-qubit gate. The magnitude of this error is comparable to the single-move error in reconfigurable arrays and is currently below $10^{-3}$ \cite{Manetsch_2025_Tweezer}. It can be further reduced by decreasing the messenger qubits’ acceleration in the loading zone. Although this would increase the messenger-qubit preparation time, it would not necessarily affect the overall logical gate time, since messenger preparation can be scheduled and performed in advance.

Another constraint of reconfigurable arrays is that not all atomic moves are allowed: each row or column must move as a whole, and one row (or column) cannot pass over another~\cite{Tan_2023_Compiling}. This imposes connectivity limitations and complicates circuit compilation~\cite{Tan_2023_Compiling,wang2024atomique,10946298}. Such constraints are absent in architectures employing messenger qubits.

That said, the reconfigurable Rydberg array architecture remains the leading and most technologically mature approach to date, supported by extensive experimental validation. The architectures proposed here should therefore be regarded as prospective alternatives, whose practical competitiveness will ultimately depend on the challenges and costs associated with their experimental implementation. The concrete technological components required for such implementations are discussed in the next Section.


\subsection{Technological stack: established, emerging, and prospective components \label{sec:stack}}

Different technological components required to implement the proposed architectures are currently at various stages of maturity. Below, we discuss the present state of the art and realistic targets for these components.

\subsubsection{Multi-species layouts}

The separation of qubits into computational and messenger ones naturally motivates the consideration of multi-species atomic setups~\cite{Brickman_2009_Ultracold,Beterov_2015_Rydberg,  Zeng_2017_Entangling,Singh_2022_Dual-Element,Sheng_2022_Defect-Free,Singh_2023_Mid-circuit,Fang_2025_Interleaved,zhang2025dualtype}. When computational and messenger qubits are realized with different atomic species, this can facilitate high-fidelity two-qubit gates between a messenger and a target computational qubit, while reducing crosstalk to nearby qubits \cite{Beterov_2015_Rydberg}. Using distinct species may also be advantageous because the optimal requirements for messenger and computational qubits differ due to their distinct roles. Moreover, computational qubits themselves can be implemented using two species arranged in a checkerboard pattern, which further suppresses crosstalk \cite{Beterov_2015_Rydberg}. This, in turn, allows an increased Rydberg interaction radius and higher  messenger-qubit velocities, or, alternatively, enables reduced spacing between computational qubits.

\subsubsection{Two-qubit gates}

At present, experimental implementations of a neutral atom quantum processor typically employ the \textsf{CZ} two-qubit gate \cite{Bluvstein_2022_Quantum, reichardt2024fault,Radnaev_2025_Universal,rines2025demonstration}, with record errors approaching $10^{-3}$ and a clear trend toward further improvement \cite{Theis_2016_High-fidelity,Madjarov_2020_High,Radnaev_2025_Universal,Muniz_2025_High-Fidelity,Peper_2025_Spectroscopy,Finkelstein_2024_Universal,Tsai_2025_Benchmarking,Bluvstein_2024_Logical, reichardt2024fault,rines2025demonstration}. However, in our approach another native two-qubit gate — the \textsf{SWAP} gate — is highly desirable. This gate has received relatively little experimental attention, perhaps because it is not an entangling operation, or because the prevailing reconfigurable atom array paradigm implements \textsf{SWAP} through physical atom motion. In contrast, our architectures are designed to keep computational qubits stationary. While a \textsf{SWAP} operation can be synthesized from three \textsf{CZ} and six single-qubit Hadamard gates, such decomposition introduces a significant overhead in gate count and, consequently, a substantial reduction in logical fidelity. We therefore emphasize the importance of direct experimental implementation of the \textsf{SWAP} gate with errors comparable to those of the \textsf{CZ} gate. Theoretical analyses \cite{Schuch_2003_Natural,Wu_2012_Quantum,Ni_2018_Dipolar,Wu_2022_Unselective,Xiao_2024_Effective,Wu_2024_Soft-controlled,Sun_2024_Holonomic,Li_2024_High-tolerance,Wang_2025_Design,Wang_2025_Deterministic} indicate that this should be well possible.

Furthermore, most current realizations rely on globally addressed two-qubit gates \cite{Bluvstein_2022_Quantum, reichardt2024fault,rines2025demonstration}, whereas our approach requires individually addressed ones.  A significant step in
this direction has been recently demonstrated in Ref.~\cite{Radnaev_2025_Universal}, with individually addressed  two-qubit gates implemented in a large reconfigurable atomic array with  error well below $10^{-2}$.

\subsubsection{Single-qubit gates}

Single-qubit gates are now routinely implemented in experiments. Nevertheless, a substantial fidelity gap remains between global operations (with reported errors of about $5\times10^{-5}$~\cite{Sheng_2018_High-Fidelity}) and individually addressed ones (recently achieving $\sim2\times10^{-4}$~\cite{Radnaev_2025_Universal}). Because individual addressability is essential for our architecture, further improvements in the fidelity of individually addressed single-qubit gates \cite{Labuhn_2014_Single-atom,Beterov_2021_Implementation-1,Beterov_2021_Implementation-2,Jenkins_2022_Ytterbium,Radnaev_2025_Universal} will directly benefit its realization.

\subsubsection{Qubit readout}

Two of the proposed architectures — the one-way conveyor belt and throw-and-measure designs — rely on mid-circuit readout of messenger qubits. Currently, the most widely used detection method is fluorescence imaging, whose relatively long readout times (typically thousands of microseconds) preclude mid-circuit operation \cite{schmid2024computational,Wintersperger_2023_Neutral}. However, fast readout schemes exploiting qubit coupling to optical cavities or ensembles of ancilla atoms are actively being developed \cite{Bochmann_2010_Lossless,volz2011measurement,Xu_2021_Fast,Deist_2022_Mid-Circuit,Petrosyan_2024_Fast,Grinkemeyer_2025_Error-detected}. These approaches promise readout times in the microsecond range, comparable to gate times and thus compatible with the requirements of our architectures. Implementation will be further facilitated by the fact that readout occurs outside the computational zone, thereby avoiding unwanted disturbance of computational qubits.

\subsubsection{Moving qubits: loading, shuttling, shooting, and routing}

The proposed architectures rely critically on a versatile toolbox for coherent control of atomic motion. Fortunately, such techniques are rapidly advancing—primarily in the context of reconfigurable atom arrays—and many relevant capabilities already exist, especially for shuttling atoms with optical tweezers. Below we summarize the key elements.

First, all proposed architectures require the capability to continuously load atoms from a reservoir into moving optical tweezers, lattices, or arrays. This capability, an essential prerequisite for scalable devices, has been recently demonstrated in several experiments~\cite{Pause_2023_Reservoir-based,Norcia_2024_Iterative,Gyger_2024_Continuous,Pause_2024_Supercharged,chiu2025continuous,li2025fast}.

The transfer of atoms via optical conveyor belts—central to three of our proposed architectures—has also been experimentally demonstrated with excellent control and very low loss rates ~\cite{Bluvstein_2022_Quantum, Evered_2023_High-fidelity,Bluvstein_2024_Logical,reichardt2024fault,Radnaev_2025_Universal}.

The two architectures that do not employ conveyor belts, namely throw-catch-throw and throw-and-measure, instead rely on messenger qubits flying freely through the computational zone. This requires precise launching of atoms along well-defined trajectories with a controlled motional wavepacket spreading. This capability has been demonstrated, though currently only for a short displacements of a few micrometers~\cite{Hwang_2023_Optical}. Scaling this technique to hundreds or thousands of micrometers represents a major but tractable next step. In addition, the catch-and-throw architecture demands the ability to “catch” -- i.e., controllably decelerate -- freely flying atoms entering a catch zone, which has also been experimentally shown in Ref.~\cite{Hwang_2023_Optical}.

Finally, the shuttle-and-route architecture requires the transfer of atoms between moving conveyor belts. While a direct demonstration of this capability has not yet been reported, a closely related operation, the transfer between static and moving tweezer arrays, have been demonstrated in reconfigurable atom arrays \cite{Bluvstein_2022_Quantum, reichardt2024fault,Radnaev_2025_Universal,rines2025demonstration,Manetsch_2025_Tweezer}, with record fidelities reaching $2\times10^{-3}$~\cite{Manetsch_2025_Tweezer}.

In all our proposed schemes, most qubit manipulations (single- and two-qubit gates, as well as mid-circuit measurements) must be performed on moving atoms (or on a pair of atoms where one is moving and another one is at rest). This requirement poses its strongest constraints on two-qubit gates, ultimately limiting the messenger qubit velocity and hence the logical gate time, as discussed in Sec.~\ref{sec: time}. Optimized gate and measurement protocols that explicitly account for the center-of-mass motion of messenger qubits will likely be needed. Such operations remain to be demonstrated experimentally.

We finally remark that executing complex quantum circuits will benefit from employing classical precompilation to optimize the sequence of hardware operations --- a fact already recognized and addressed in the context of the reconfigurable atomic arrays  ~\cite{brandhofer2021optimal,Tan_2022_Qubit,  Tan_2023_Compiling, wang2024atomique,Stade_2024_Abstract,Lin_2025_Reuse,stade2025routing,huang2024zap}.

\subsubsection{Comparison between the five architectures}

Each of the five architectures considered here imposes distinct requirements on the technological stack, offering specific advantages and challenges (see Table~\ref{tab:Ngates} and Figs.~\ref{fig:errors} and ~\ref{fig:errors_measurement}). The optimal choice depends on which technologies are available. For example, if fast, high-fidelity mid-circuit readout of messenger qubits is achieved, the one-way conveyor belt and throw-and-measure architectures become particularly attractive due to their reduced gate count and lower cumulative gate errors. If precise launching and targeting of free-flying atoms by means of throw tweezers are realized, the throw-catch-throw and throw-and-measure architectures offer similar benefits. Demonstration of atom transfer between moving conveyor belts would open the way for the shuttle-and-route design. Conversely, if one is restricted to technologies already demonstrated experimentally, the two-way conveyor belt architecture remains the most immediately feasible option, albeit with the highest gate count and hence the most stringent requirements on gate fidelities. Finally, hybrid architectures combining elements of several designs may provide additional flexibility and optimization opportunities.


\section{Conclusions}\label{sec:conclusion}

We have proposed an approach to the neutral-atom quantum computer architecture in which entanglement between distant computational qubits is mediated by moving messenger qubits. In this way, the connectivity hurdle is resolved with a moderate, size-independent overhead in the gate count. By employing disposable messenger qubits that travel with a constant velocity most of their lifetime, heating errors incurred by the atomic movements become independent of both the processor size and the circuit depth.

We have outlined five specific architectures implementing this concept, each characterized by a distinct balance of advantages and challenges. Many of the technological components required for their realization already exist, while others are only beginning to emerge. We have identified several key elements that demand further development or refinement --- most notably, locally addressable single- and two-qubit gates (including the \textsf{SWAP} gate, which has so far received little experimental attention), and gate operations involving moving atoms, particularly in dual-species layouts. For architectures incorporating mid-circuit qubit readout, rapid measurement of messenger qubits is also essential; this requirement is eased by the fact that such readout can be performed outside the computational zone. Taken together, these developments seem attainable in the near to mid term, opening a new avenue to a fully connected neutral-atom quantum processor.

\medskip
\noindent {\it Note added.} ~~We draw the reader’s attention to the recent Ref.~\cite{saffman2025quantum}, which discusses the challenge of ensuring connectivity in neutral-atom quantum processors and highlights the accumulation of heating errors arising from multiple nonuniform atomic moves --- consistent with our analysis.

\begin{acknowledgements}
We acknowledge the support of the Russian Roadmap on Quantum Computing (Contract No. 868-1.3-15/15-2021). The work of A.K.F. is supported by the Priority 2030 program at the NIST “MISIS” under the project K1-2022-027.
\end{acknowledgements}

\bibliography{rydberg-flying.bib}

\end{document}